\documentstyle[multicol,aps,epsf,pra]{revtex}

\begin{document}
\draft

\title{ A Multiparticle Generalization of Entanglement Swapping}
\author{ S.Bose, V.Vedral  and P.L.Knight}
\address{Optics Section, The Blackett Laboratory, Imperial College, London SW7 2BZ, England}
\date{\today}

\maketitle
\begin{abstract}
We generalize the procedure of entanglement swapping to obtain a scheme for manipulating entanglement in multiparticle systems. We 
describe how this
scheme allows to  establish  multiparticle entanglement between particles belonging to distant users in a 
communication network through a prior distribution of singlets followed
by only local measurements. We show that this scheme can be regarded as a method of
generating entangled states of many particles and compare it with existing
schemes using simple quantum computational networks. We highlight the
practical advantages of using a series of entanglement swappings during
the distribution of entangled particles between two parties.
Applications of
multiparticle entangled states in cryptographic conferencing and in reading messages from more than one source through a single measurement are
 also described.  
\end{abstract}
\pacs{Pacs Nos: 03.65.Bz}

\begin{multicols}{2}

\section{Introduction}
       There are numerous uses of spatially separated entangled pairs of particles such as cryptography based on Bell's theorem \cite{Ek}, teleportation \cite{ben}, superdense coding \cite{wies}, cheating bit commitment \cite{lo} and, of course, simply
testing Bell's inequalities \cite{Bell,braun}. It is natural to expect that three or more spatially separated particles in an entangled
state (such as a GHZ state \cite{ghz}) will have varied applications as well. A well known application is in testing nonlocality from new directions \cite{ghz,mult,peres}. More recently, applications such as 
reducing communication complexity \cite{distr} and quantum telecomputation \cite{grov} have been
suggested. An idea of cryptography with GHZ states \cite{mcrypt} has also been
suggested earlier. This essentially means that multiparticle entangled states  present possibilities of implementing networked cryptographic conferencing \cite{barn,mor} in an alternative way and we shall elaborate briefly on this
aspect in section \ref{use} of this article. In addition, we shall show (by
generalizing superdense coding \cite{wies} to the multiparticle case), that the use of multiparticle entangled states can allow one to read messages from more than one source through a single measurement. In short, distributed entangled particles in a multiparticle entangled state may be extremely useful for certain forms of quantum communication. Hence it will become necessary to distribute entangled 
N-tuplets between nodes of communication networks if any set of users of the
 network wish to harness the benefits of these forms of quantum communication.  The prime focus of this paper is to present a method of manipulating entanglement in multiparticle systems that can accomplish this task by just a local measurement if each of the users share singlets with a central node prior to that. In a sense, it allows one to
construct a Biham-Huttner-Mor (BHM) like telephone exchange \cite{mor}, with the 
added capability of setting up multiparticle entanglement between particles
belonging to different users of the network. 

       Our scheme is actually obtained by generalizing an earlier scheme of Zukowski et al \cite{swap} known as {\em entanglement swapping} to the multiparticle case. Building on an earlier proposal by Yurke and Stoler \cite{yk} of entangling particles originating from independent sources,
they showed that through entanglement swapping one can entangle particles 
which do not even share any common past. Their aim was to realize 'event ready detectors' for
Bell experiments. In section \ref{prac} we point out two ways in which
their original entanglement swapping scheme can come to a practical 
advantage while distributing entangled particles between two parties. We also point out that
our form of multiparticle entanglement  manipulation differs from the method proposed recently by Zeilinger et al \cite{zein} for the generation
of multiparticle entangled states  in the application of a few quantum gates and yet can
be used for the same purpose.  We begin by briefly recapitulating the
original entanglement swapping scheme of Zukowski et al \cite{swap} in Section.\ref{swp}. 

\section{The entanglement swapping scheme of Zukowski et al}
\label{swp}
    In terms of a binary variable  $u_i \in \{0,1\}$ and its complement $ u_i^c $
(defined as $1-u_i$),  one can write down any Bell state (not normalized) of  two particles $i$ and
$j$ as 
\begin{equation}
\label{b1}
     |\Psi(i,j)\rangle_{\pm} = |u_i, u_j \rangle \pm |u_i^c, u_j ^c\rangle ~.
\end{equation}
In the above it is understood that $|u_i\rangle$ and $|u_i^c\rangle$ are  two
orthogonal states of a two state system.  Consider the initial state of four
particles 1,2,3 and 4 to be 
\begin{equation}
\label{ini}
\begin{array}{rcrcl}
    |\Psi(1,2)\rangle_+  \otimes  |\Psi(3,4)\rangle_+ & = &   
   |u_1,u_2,u_3,u_4 \rangle  \nonumber \\
& + & |u_1^c,u_2^c,u_3,u_4 \rangle    \nonumber  \\
& + & |u_1,u_2,u_3^c,u_4 ^c\rangle \nonumber \\
& + & |u_1^c,u_2^c,u_3^c,u_4^c \rangle
  ~.
\end{array}
\end{equation}
That is, particles 1 and 2 are mutually entangled (in a Bell state), and particles 3 and 4 are mutually entangled (also in a Bell state). 
 When we conduct a measurement of the Bell operator (defined in \cite{braun})
on particles 2 and 3 (which projects particles 2 and 3 to a Bell state), then the 
joint state of the four particles become either of the following four :
\begin{mathletters}
\begin{equation}
\label{bl1}
 |\Phi_1 \rangle =    (|u_2, u_3 \rangle + |u_2^c, u_3^c \rangle) \otimes  (|u_1, u_4 \rangle + |u_1^c, u_4^c \rangle)  ~,\end{equation}
\begin{equation}
\label{bl2} 
 |\Phi_2 \rangle =   (|u_2, u_3 \rangle - |u_2^c, u_3^c \rangle) \otimes
   (|u_1, u_4 \rangle - |u_1^c, u_4^c \rangle)    ~, \end{equation}

\begin{equation}
\label{bl3} 
 |\Phi_3 \rangle =   (|u_2, u_3^c \rangle + |u_2^c,  u_3\rangle) \otimes      
    (|u_1, u_4^c  \rangle + |u_1^c, u_4\rangle)  ~,  \end{equation}

\begin{equation}
\label{bl4} 
 |\Phi_4 \rangle =   (|u_2, u_3^c\rangle - |u_2^c,  u_3 \rangle) \otimes     
    (|u_1, u_4^c \rangle - |u_1^c, u_4 \rangle)  ~. \end{equation}
\end{mathletters}

To derive the above, only the orthogonality of $|u_i\rangle$ and $|u_i^c\rangle$
is required.
In other words, no matter what the outcome is, the particles 1 and 4 are
now in one of the Bell states . Whereas prior to the measurement, the Bell
 pairs were (1,2) and (3,4), after the measurement the Bell pairs are (2,3)
and (1,4). It can easily be shown that the same fact would hold true even
if (1,2) and (3,4) started in some other Bell states than those in Eq.(\ref{ini}). A pictorial way of representing the above process is given in Fig.1. 

\vspace{1cm} 
\begin{figure} 
\begin{center} 
\leavevmode 
\epsfxsize=7cm 
\epsfbox{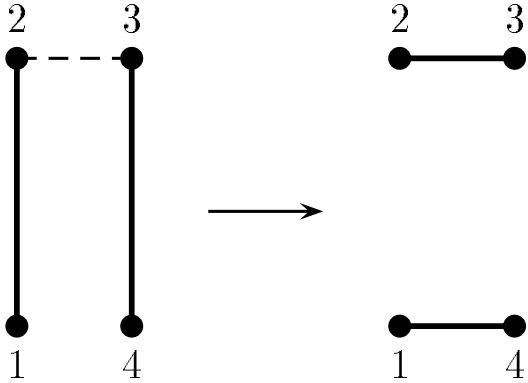}
\vspace{1cm}
\caption{\narrowtext The swapping of entanglement between pairs of particles due to a Bell state measurement on two of them is shown. The bold lines connect particles
in Bell states and the dashed lines connect particles on which the Bell state
measurement is made}
\end{center} 
\end{figure}
\vspace{1cm}

     It is clear that the most interesting aspect of this scheme is that
particles 1 and 4 which do not share any common past are entangled after
the swapping.

\section{A multiparticle generalization of entanglement swapping}
\label{man}
 
The method of entanglement manipulation described in the previous section
 can readily be generalized to cases where a greater number of particles are involved. An explicit scheme that generalizes entanglement swapping to the
case of generating a 3-particle GHZ state from three Bell pairs has already
been presented by Zukowski et al \cite{zk2}. But it should be interesting
to demonstrate that their entanglement swapping scheme may actually be significantly generalized
to the case of starting with cat states involving any number of particles,
doing local measurements by selecting any number of particles from the different cat states and also ending up with cat states involving any number of particles ( The cat state is a terminology
used to refer to generalizations of Bell states and GHZ states to higher number
of particles).  To see that consider an initial state in which there are $N$ different sets of entangled particles in cat states. Let each of these sets be labelled by
$m$ (where $m=1,2,..,N$), the $i$th particle of the $m$th set be labelled by
$i(m)$ and the total number of particles in the $m$th set be $n_m$. Then the
initial state can be represented by
\begin{equation}
\label{Nini}
        |\Psi\rangle = \prod_{m=1}^N |\Psi\rangle_m   ,
\end{equation}
in which each of the cat states $|\Psi\rangle_m $ is given by
\begin{equation}
\label{mini}
        |\Psi\rangle_m = \prod_{i=1}^{n_m} |u_{i(m)} \rangle \pm \prod_{i=1}^{n_m} |u_{i(m)}^c \rangle 
\end{equation}
where the symbols
$u_{i(m)}$ stand for  binary variables $\in \{0,1\}$ with $u_{i(m)}^c=1-u_{i(m)}$. Now imagine that the first $p_m$ particles from all the  entangled sets are brought together (i.e a total of $p=\sum_{m=1}^N p_m$ particles) and a joint measurement is performed on all of them. Note
that the set of all cat states of $p$ particles forms a complete
orthonormal basis. Let the nature of the measurement on the selected particles be such that
it projects them to this basis. Such a basis will be composed of states of the type
\begin{equation}
\label{e1fin}
 |\Psi(p)\rangle = \prod_{m=1}^N \prod_{i=1}^{p_m} |u_{i(m)} \rangle \pm \prod_{m=1}^N \prod_{i=1}^{p_m} |u_{i(m)}^c \rangle    .
\end{equation}
By simply operating with $|\Psi(p)\rangle \langle \Psi(p)|$  on $|\Psi\rangle $ of Eq.(\ref{Nini}), we find that the rest of the particles (i.e those not being measured)
are projected to states of the type
\begin{equation}
\label{e2fin}
 |\Psi(\sum_{m=1}^N n_m-p)\rangle = \prod_{m=1}^N \prod_{i=p_m+1}^{n_m} |u_{i(m)} \rangle \pm \prod_{m=1}^N \prod_{i=p_m+1}^{n_m} |u_{i(m)}^c \rangle    ,
\end{equation}
which represents a cat state of the rest of the particles. In a
schematic way the above process can be represented as
\begin{equation}
   \prod_{m=1}^N |E(n_m)\rangle \rightarrow |E(p)\rangle \otimes |E(\sum_{m=1}^N n_m-p)\rangle
\end{equation}
where $|E(n)\rangle$ denotes a $n$ particle cat state. As a specific
example, in Fig.2, we have shown the conversion of a collection of two Bell states and a 3 particle GHZ state to a 3 particle GHZ state and a 4 particle GHZ state
due to a projection of 3 of these particles to a 3 particle GHZ state.

\vspace{2cm} 

\begin{figure} 
\begin{center} 
\leavevmode 
\epsfxsize=7cm 
\epsfbox{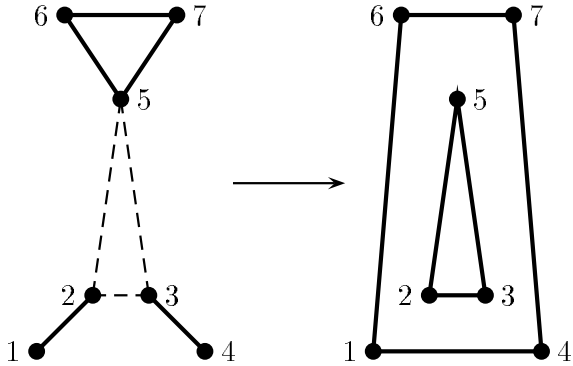}

\vspace{2cm}

\caption{\narrowtext The conversion of two Bell states and a 3 particle GHZ state to a 3 particle GHZ state and
a 4 particle GHZ state due to a GHZ state projection on three
particles (one taken from each of the initially entangled sets) is shown. The bold lines connect mutually entangled particles and the dashed lines connect particles on which the GHZ state
projection is made.}

\end{center} 
\end{figure}

 As must be evident from Fig.2, there is a general pencil and paper rule
for finding out the result when our method of entanglement  manipulation is applied to a certain collection of cat states of particles. One just has to connect the particles being measured to frame a polygon and those not being measured to frame a complementary polygon. These two polygons represent the
two multiparticle cat states obtained after the manipulation.

\section{Establishing multiparticle entanglement between particles located at different nodes
  of a communication network} We now describe how our method of multiparticle entanglement manipulation can be used to set up entanglement between
particles belonging to N users in a communication network.
To begin with, each user of the network needs to share entangled pairs of particles
(in a Bell state) with a central exchange. Consider Fig.3:  A, B, C and D are users who  share the Bell pairs (1,2), (3,4), (5,6) and (7,8) respectively with a central exchange O. Now suppose that A, B and C wish to share a GHZ triplet. Then a measurement which projects particles
2, 3 and 5 to GHZ states will have to be performed at O.  Immediately, particles 1, 4 and
6 belonging to A, B and C respectively will be reduced to a GHZ state. In a 
similar manner one can entangle particles belonging to any N users of the network and create a N particle cat state. 
          
         The main advantages of using this technique for establishing entanglement over the simple generation of N particle entangled states
at a source and their subsequent distribution are as follows. 

      (A) Firstly, each user can at first purify \cite{pur} a large number of partially decohered Bell pairs shared with the
central exchange to obtain a smaller number of pure shared Bell pairs. These
can then be used as the starting point for the generation of any types of multiparticle
cat states of the particles possessed by the users. The problems of decoherence during propagation of the particles can thus be avoided in principle. Also the necessity of having to purify N-particle cat states
can be totally evaded. Purification of singlets followed by our scheme will
generate N-particle cats in their purest form. 

     (B) Secondly, our method allows
a certain degree of freedom to entangle particles belonging to any set of users
only if the necessity arises. It may not be known in advance exactly which set
of users will need to share a N particle cat state. To arrange for all
possibilities in an a priori fashion would require selecting
all possible combinations of users and distributing particles in  multiparticle entangled states among them. That is very uneconomical. On the other
hand, generating entangled N-tuplets at the time of need
 and supplying them to the users who wish to communicate is definitely time consuming.

\vspace{1cm} 

\begin{figure} 
\begin{center} 
\leavevmode 
\epsfxsize=7cm 
\epsfbox{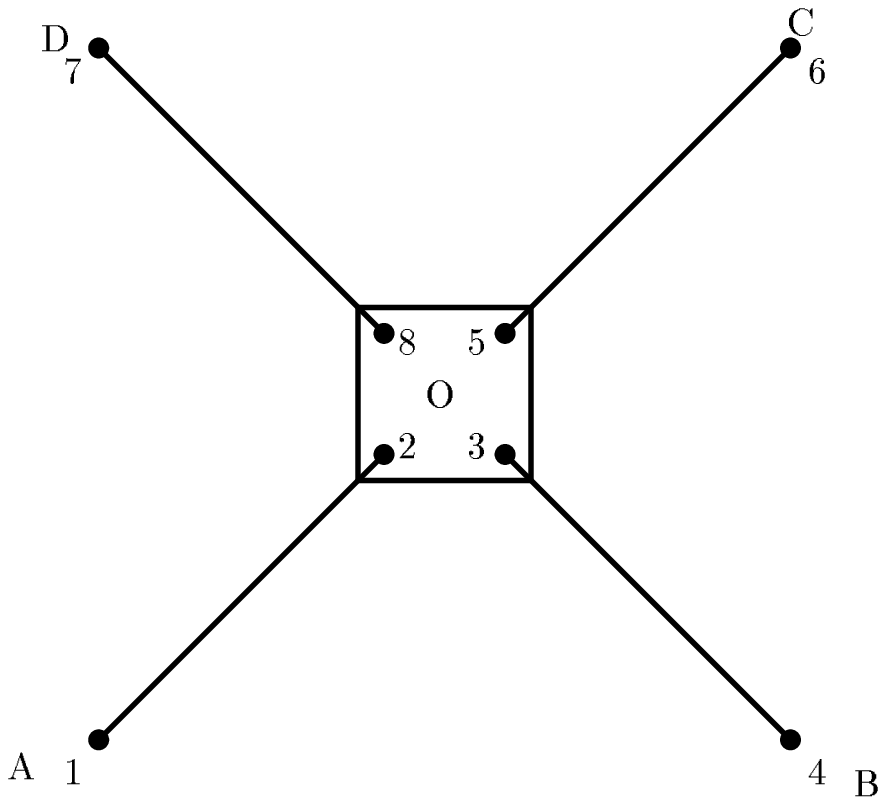}

\vspace{2cm}

\caption{\narrowtext The configuration used for the distribution of entanglement. Initially users
A,B,C and D share Bell pairs with the central exchange O. Subsequently, a local
measurement at O is sufficient to entangle particles belonging to any
subset of users chosen from A, B, C and D .}

\label{setup}
\end{center} 
\end{figure}

\vspace{1cm}

      It is pertinent to compare our scheme with the BHM cryptographic network with exchanges \cite{mor}. There are two main differences. 

 (A) Firstly, they used a time reversed EPR scheme for setting up the connections and had quantum memories to protect their states. They had, of course, compared their scheme with one that uses the original entanglement swapping of two Bell pairs
to establish connections. We,  use a multiparticle generalization of entanglement
swapping for our connections and are thereby unable to take advantage of any type of quantum memories.

 (B) Secondly, their prime focus was to connect any pair of users of a N-user network on
request, while our main focus is to establish multiparticle entangled states
of the particles possesed by the users.

\section{Practical uses of standard entanglement swapping}
\label{prac}
\subsection{Speeding up the distribution of entanglement}          
     We now explain how standard entanglement swapping helps to save a significant amount of
time when one wants to supply two distant users with a pair of atoms or electrons (or any particle possesing mass) in a Bell state
from some
central source. The trick is to place several Bell state producing and  Bell state measuring substations in the
route between them. Consider Fig.4(a):  A and B are two users separated by a distance
L; O, which is situated midway between A and B is a source of Bell pairs. The time
needed for the particles to reach A and B is at least $t_1=L/2v$ where $v < c$  (the speed of light) is the speed of the particles. Now consider Fig.4(b)
in which two  Bell pair producing stations C and D are introduced halfway between
AO and BO respectively and O is now just a Bell state measuring station. At $t=0$, both C and D send off Bell pairs (1,2) and (3,4) respectively. 2 and 3 arrive at O, 1 reaches A and 4 reaches B. They all arrive at their destinations  exactly at $t= L/4v$.
At this instant a Bell state measurement is performed on  particles 2 and 3
 at O. This measurement immediately
reduces the particles 1 and 4 reaching A and B respectively, to a Bell state.  If the time of measurement
is denoted by $t_m$, then the time needed to supply a Bell pair to A and B with the 
two extra substations C and D on the path is  $ t_2 = L/4v + t_m $. It is 
evident that $t_2$ is lesser  than $t_1$ if $t_m < L/4v$. Of course, to this time one needs to add the time needed to do classical communication
between the station O and the users A and B to communicate the particular
Bell state to which particles 1 and 4 are projected.  So for photons in
Bell states, this procedure cannot really save any time. But for particles
possesing mass, this is definitely a way to reduce the time needed to supply
to distant users with a Bell pair. In this way one
can reduce the time needed to supply two distant users with a Bell pair
even further by including more and more Bell pair producing and measuring substations on the way.

\begin{figure} 
\begin{center} 
\leavevmode 
\epsfxsize=7cm 
\epsfbox{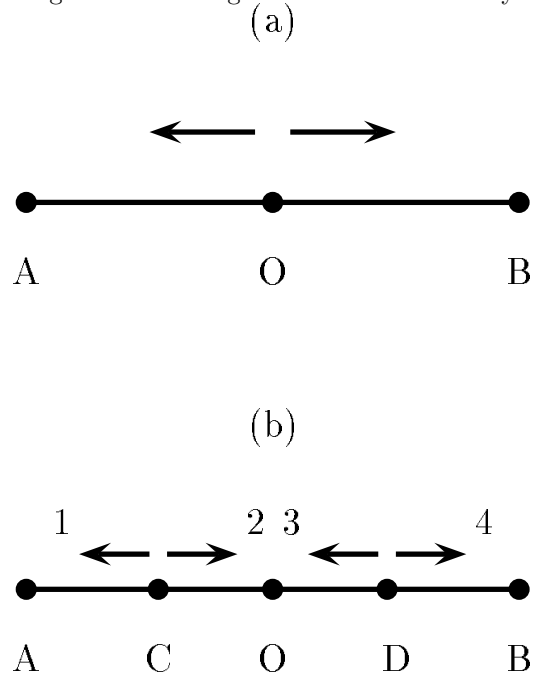}

\vspace{2cm}

\caption{\narrowtext A method of increasing the speed of distributing an entangled pair
of particles (that posses mass) between two distant users A and B is illustrated. Extra Bell state generating substations C and D are inserted between A and B and a Bell state
projection is performed at O to speed up the distribution of a Bell pair between
A and B.}

\label{setup}
\end{center} 
\end{figure}

\subsection{Correction of amplitude errors developed due to propagation}
                      We would like to show that entanglement swapping can be used, with some probability which we quantify, 
to correct amplitude errors that might develop in maximally entangled states
during propagation. Assume that in Fig.4(b), the Bell pairs emitted from
C and D acquire amplitude errors and become less entangled states of the type
   \begin{equation}
\label{ampl}
       |\Psi\rangle = \cos{\theta}|01\rangle + \sin{\theta}|10\rangle .
  \end{equation}
Thus, the combined state of the two entangled pairs, when particles 2 and 3
reach O is given by,
\begin{eqnarray}
|\Phi\rangle & = & \cos^2{\theta}|0101\rangle + \sin{\theta}\cos{\theta}(|1001\rangle \nonumber \\ 
& + & |0110\rangle) + \sin^2{\theta}|1010\rangle. 
\end{eqnarray}

If a Bell state measurement is now performed on particles 2 and 3 that reach
O, then the probability of them being projected onto the Bell states 
$|00\rangle+|11\rangle $ or $|00\rangle-|11\rangle $ is $\sin^2{2\theta}/2$, while the probability
of them being projected onto any of the other two Bell states is $(1+\cos^2{2\theta})/2$. In the first case (i.e when 2 and 3 get projected
to $|00\rangle+|11\rangle$ or $|00\rangle-|11\rangle $ ), the distant particles 1 and 4 are projected
onto the Bell states $|00\rangle+|11\rangle$ or $|00\rangle-|11\rangle $. In this way in spite of amplitude
errors due to propagation of the particles, A and B may finally share a Bell state. Of course in case of the other two outcomes of the state of particles 2 and 3, particles 1 and 4 go to states even less entangled than that of
Eq.(\ref{ampl}). That is why we can consider entanglement swapping suitable for correction
of amplitude errors only probabilistically. The probability of success in
this case ($\sin^2{2\theta}/2$), is lower than the probability of
failure ($(1+\cos^2{2\theta})/2$). However, from the outcome of the
Bell state measurement, one knows when the correction has been successful.
This may be regarded as a kind of purification in series in contrast to
the standard purifications \cite{pur} which occur in parallel. 
      
    It should be noted that earlier there have been suggestions \cite{div} of 
placing several stations in series in the path between two distant users, purifying
singlets shared by adjacent stations and then using teleportation from one station to the next  to derive purified singlets shared by the two users.
The suggestion here is quite different in the sense that the methodology
of entanglement swapping itself is being used for supplying a pair of distant users with a pure Bell pair and no separate purification procedure is invoked.

\section{ Communication schemes using distributed multiparticle entanglement}
\label{use}    

   So far we have described how particles belonging to several users in a network can be put in a multiparticle entangled state. We now describe how such a situation can be useful. The  uses that readily come to mind are tests of nonlocality  for many particle entangled states
\cite{ghz,mult,peres}. Reducing the communication complexity of certain functions \cite{distr} may also be a field of
application of multiparticle entangled states. We describe here two possible applications
in communications.
\subsection{Cryptographic conferencing}  
     Consider the following situation. A certain group of users may need to have a secret meeting. Let the meeting be subject to the following two conditions: 

             (1) All members of the
group must be able to decrypt the encrypted public messages broadcasted by any  member of the group.
   
           (2) Nobody outside the group must be able to decrypt the encrypted public messages broadcasted by members of this group. 

         Such requirements may arise when the members of a committee are to take some decision which has to have the
consent of everybody within the committee, but must be kept secret from the rest of the world. One can regard this as a special
case of networked {\em cryptographic conferencing} \cite{barn,mor}.To establish the secret key for
this type of meeting, the committee can employ either of two possible options. The first one is to use
simple two-user cryptographic key distributions \cite{Ek,crypt}, to set up random keys shared
by each pair of users. Whenever a certain user intends to broadcast a secret message for
the group, she has to encrypt it using a separate key for each of the other users of the
group. However, for the type of conferencing
considered here, this is not a good option because of two reasons. Firstly, the broadcaster
has to encrypt the same message several times using different keys. Secondly, the broadcaster
will have the freedom to send different messages to different members of the group and thereby
mislead a subset of the group. A better option is that the users within the group share particles in multiparticle maximally entangled states. To generate the random key,
all users conduct measurements in two nonorthogonal bases on the particles belonging to
them. The results of those measurements in which the bases chosen by all the users 
coincide, are used to establish the secret key known to all the users within the group.
This can then be used to frame encrypted messages that can be decrypted
by an user if and only if she is a member of the group. If the users are
sure about the fact that they are sharing a perfect N-particle cat state,
they can even use a single basis to perform their measurements and thereby
reduce the wastage of bits due to the noncoincidence of all their bases.
  It is known that the 3 particle GHZ states are eigenstates of certain
operators of the form $ S_x S_x S_y $ \cite{ghz} and by measuring the eigenvalues
of these operators one can verify whether the state is intact or corrupted
by some evesdropper. But measuring these operators should not change the particular
GHZ state in which the particles are because this state is an eigenstate of the
operator being measured. So, {\em in principle}, while doing a 3 party
cryptography with a 3 particle GHZ state, one can essentially use the same set
of particles for verifying the fidelity of the GHZ state and the susequent
establishment of a secret key.   It should be mentioned here that the idea
of using GHZ states for cryptography is not entirely new but has been 
presented earlier \cite{mcrypt}.

\subsection{A multiparticle generalization of entanglement swapping}
             Sharing particles in a multiparticle entangled state may also help an user
to read messages from more than one user through a single measurement. This
is a kind of generalization of the well known superdense coding scheme \cite{wies}. Suppose $N+1$ users are sharing a $N+1$ particle maximally
entangled state, possessing one particle each. Also suppose that one of them,
say user 1, intends to receive messages from the $N$ other users, whom we shall
refer to as senders. This can be done in the following way.  The $N$ senders will have to mutually decide a priori
to perform only certain unitary operations on the particles given to them. One
of the senders will have any of 4 possible unitary operations at his/her disposal,
while each of the others will have any of 2 possible unitary operations at
their disposal. They will be encoding bits onto the particles possessed
by them through these unitary operations. In that way, one of the senders
will be encoding 2 bits on his/her particle, while each of the others
will be encoding 1 bit each. Now, the unitary transformations must be
so chosen that for each possible combination of unitary transformations performed by
the senders, the state of the $N+1$ particles go to a different member of the set
of maximally entangled states of $N+1$ particles. This is certainly possible     because the $N$ senders are allowed to perform $4\times2\times2\times...\times2 = 2^{N+1}$ different combinations of unitary operations on the initial
$N+1$ particle maximally entangled state and there are exactly $2^{N+1}$ 
states in the set
of maximally entangled states of $N+1$ particles. After performing
their unitary operations, each of the $N$ senders send their particles to
user 1. User 1 then performs a measurement on the $N+1$ particles possessed
by him, which identifies the particular maximally entangled state in which these
particles are. Since each of these maximally entangled states correspond
to a different combination of unitary transformations performed by the
senders, he can learn about the messages sent by each of the senders
from the outcome of his measurement. Thus a single measurement is sufficient
to reveal the messages sent by more than one user.

\begin{figure} 
\begin{center} 
\leavevmode 
\epsfxsize=10cm 
\epsfbox{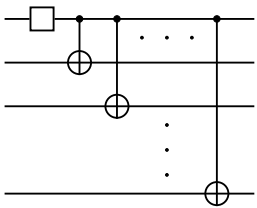}

\caption{\narrowtext The general circuit to generate a $N$ particle maximally entangled state
from disentangled inputs is shown. One can generate any desired state chosen from the basis of maximally entangled states of $N$ particles by inputting
an appropriate combination of zeros and ones
 at the input port on the left hand side. When we input any specific $N$ particle maximally entangled state from the right hand side, then by measuring the
bits obtained on the left hand side, one can conclude which maximally entangled state was inputted. In other words, when run in the reverse direction, the
circuit represents a measuring device for $N$ particle maximally entangled states. The box in the figure stands for a Hadamard transformation . }

\label{setup}
\end{center} 
\end{figure}

          Now let us compare the efficiency of the above scheme with the
case when user 1 performs a superdense coded communication \cite{wies}
with each of the $N$ senders separately. For that we have to refer to
the quantum circuits needed to perform measurements of multiparticle
maximally entangled states. The circuit shown in Fig.5 does exactly this
when run in the reverse direction (i.e inputs from the right and output
from the left). The vertical lines are Controlled Nots \cite{Barenco}
and the box is a unitary transformation equivalent to a rotation on the Bloch
sphere known as the Hadamard transformation \cite{Barenco} (For optical implementations of elementary gates, see e.g. \cite{kw}). Thus, if user 1
was using our scheme then she would need to measure a $N+1$ particle
maximally entangled state and therefore require $N$ Controlled Nots and
one Hadamard transformation gate. She learns $N+1$ bits of information from
this measurement. Thus if a Hadamard transformation takes time $t_h$ and 
a Controlled Not takes time $t_c$, then the rate of information gain is
\begin{equation}
       r_1= \frac{N+1}{t_h + N t_c} 
\end{equation}
bits per unit time. On the other hand, if user 1
was separately doing superdense coded communication with each of the $N$ senders, she would need to do $N$ Bell state measurements which require
one Controlled Not and one Hadamard transformation each. She gains $2N$ bits
of information from this. Thus the rate of information gain in this case is 
\begin{equation}
       r_2= \frac{2N}{N(t_h + t_c)} .
\end{equation}
If one assumes that quantum gates have the same time scale of operation (that
is, $t_h=t_c$), then $r_2$ is exactly equal to $r_1$. However if one
defines efficiency as rate divided by the number of particles used (considering particles and channels required to propagate them to be important resources), then our method is definitely more efficient
because it requires only $N+1$ particles as opposed to superdense coded
communication with each sender which requires $2N$ particles.
   
\section{Entangled states of higher number of particles from entangled states of lower number of particles} 
      Entangled states involving higher number of particles can be generated from 
entangled states involving lower number of particles by employing our scheme. The
basic ingredients which we need  are  GHZ (three particle maximally
entangled) states and a Bell state measuring device. Let us describe how to proceed
from a $N$ particle maximally entangled state to a $N+1$ particle maximally entangled
state. One has to take one particle from the $N$ particle maximally entangled state
and another particle from a GHZ state and perform a Bell state measurement on these
two particles. The result will be to put these two particles in a Bell state  and
the remaining $N+1$ particles in a maximally entangled state . Symbolically, the way
of proceeding from a $N$ particle maximally entangled state to a $N+1$ particle maximally
entangled state is given by
\begin{eqnarray}
|E(N)\rangle \otimes |E(3)\rangle \stackrel{\mbox{Bell State Meas.}}{\longrightarrow} |E(N+1)\rangle \otimes |E(2)\rangle. \nonumber
\end{eqnarray}
An example of proceeding from a 4 particle maximally entangled state to a 5 particle maximally
entangled state by the above procedure is shown in Fig.5 . 

\vspace{2cm} 

\begin{figure} 
\begin{center} 
\leavevmode 
\epsfxsize=8cm 
\epsfbox{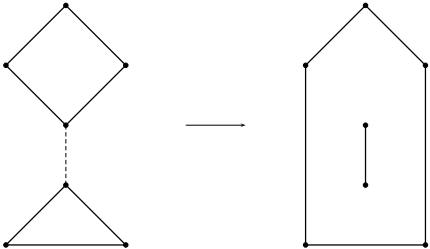}

\vspace{2cm}

\caption{\narrowtext Building of a 5 particle entangled state from a 4 particle entangled state
using a GHZ state and a Bell state measurement}

\label{setup}
\end{center} 
\end{figure}

\vspace{1cm}

As far as the question of generating the GHZ state, which is a basic ingredient, is concerned, one can perhaps use the method suggested by Zeilinger et al. \cite{zein} which is described in the next section. Alternatively, one can generate
GHZ states using our method by starting from three Bell pairs and performing a GHZ state measurement taking one particle from each pair. An explicit
scheme to produce 3-particle GHZ states from 3 entangled pairs have
been suggested by Zukowski et al. \cite{zk2} earlier.

\section{Comparison with the method of Zeilinger et al.}

We would now like to compare our method of generating multiparticle
entangled states to that of Zeilinger et al \cite{zein}. 
We describe how the Zeilinger group's scheme works in the spirit 
of the former sections, i.e. using projections onto the Bell states.
We then use simple quantum computational networks, consisting of 
Controlled Not and Not operations \cite{Barenco}, to present 
measurements in the Bell and generalized Bell basis. 
This quantum computational approach, 
as we will see shortly, immediately
reveals a basic similarity between the two schemes of generating
multiparticle entangled states and shows precisely how they differ.  

Assume that we start with two entangled pairs $(1,2)$ and $(3,4)$
of two-level systems in a state

\begin{eqnarray}
|\Psi (1,2)\rangle \otimes |\Psi (3,4)\rangle & = & (|0,0\rangle
+ |1,1\rangle)\otimes (|0,0\rangle + |1,1\rangle) \nonumber \\
& = &  |0,0,0,0\rangle + |0,0,1,1\rangle \nonumber \\
& + & |1,1,0,0\rangle + |1,1,1,1\rangle \;\; .
\end{eqnarray}
As before we omit the normalization. The fact that we use this particular,
``computational" basis $(0,1)$ does not limit the generality of our
method, but is more convenient in describing the action of quantum gates.
Suppose now that we affect a Controlled Not gate between qubits $2$ (control)
and $3$ (target), i.e. qubit $3$ changes its value only if qubit $2$ is
in the state $|1\rangle$. The above state then becomes:
\begin{equation}
|0,0,0,0\rangle + |0,0,1,1\rangle + |1,1,1,0\rangle + |1,1,0,1\rangle
\end{equation} 
Now we measure the value of the third qubit. We have two possibilities
depending on whether the outcome of this measurement is $0$ or $1$:

\begin{eqnarray}
|0,0,0\rangle + |1,1,1\rangle & \qquad\qquad &   \mbox{outcome} \qquad 0     \\
|0,0,1\rangle + |1,1,0\rangle & \qquad\qquad &   \mbox{outcome} \qquad 1
\end{eqnarray}
 So, by the action of a 
single Controlled not gate and a measurement ,
we can create a GHZ state out of two pairs initially in a Bell state. A network
for this operation is given in Fig. 7. 

Now this directly generalizes to more entangled particles. Suppose
that we have a group of $N$ entangled particles and $M$ entangled 
particles, each one in a maximally entangled state. It is then 
enough to perform a Controlled Not operation between a particle
from the first and a particle from the second group, and then a measurement 
of the target particle. The end product will clearly be a maximally
entangled state of $N+M-1$ particles and a single particle disentangled
from the rest. This completes our description of
Zeilinger et al's scheme.

\begin{figure} 
\begin{center} 
\leavevmode 
\epsfxsize=15cm 
\epsfbox{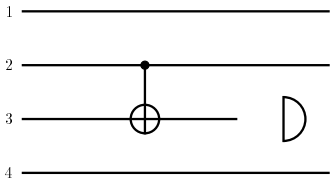}

\vspace{0.5cm}

\caption{\narrowtext Implementation of Zeilinger's scheme using quantum gates. The production of a GHZ state with the use of a controlled not, and a measurement
is shown.}

\label{setup}
\end{center} 
\end{figure}

Let us now compare this to our scheme of generating multiparticle 
entanglement. The basic ingredient in our protocol is a projection
onto a maximally entangled state of $N$ particles. The network for
a measurement in this basis is an inverse of the network which
generates maximally entangled state from a distentangled input in
the computational basis. It is simply the circuit of Fig. 5 run in the reverse
direction. In order to generate a $N+1$ particle maximally entangled state , we have to do a Bell state
measurement selecting one particle from the GHZ state and another particle from a $N$ particle maximally entangled state. A Bell state measurement will require
a Controlled Not gate and a Hadamard transformation gate, as is evident from
Fig. 5 , specializing to the case of two inputs. Meanwhile, to generate a $N+1$ particle maximally entangled state using Zeilinger group's scheme one may
require to start with a $N$ particle maximally entangled state and a Bell state
and will have to perform a Controlled Not choosing one particle from each
set followed by a measurement on one of the particles. Hence as far a 
comparing our scheme with Zeilinger group's scheme is concerned, in the case
of generating $N+1$ particle maximally entangled state from a $N$ particle
maximally entangled state, our scheme has just one extra Hadamard transformation. However, our method necessarily requires a GHZ state for
proceeding from $N$ to $N+1$ particle maximally entangled state, while
for the scheme of Zeilinger et al, Bell states are sufficient. In that sense the latter method is simpler. However
as far as setting up entanglement between particles belonging to distant
users is concerned, our method is necessary. For example, if the operator
at O in Fig.3 performed the manipulation described by the Zeilinger group manipulation on particles 2 and 3,
then 1, 2 and 4 will go into a GHZ state and essentially the subsystem
of particles 1 and 4 will be in a disentangled state. This method will thus
not be applicable to setting up entanglement between particles belonging to distant
users.

\section{Conclusion}

In this paper we have described a scheme for manipulation of multiparticle
entangled states and have demonstrated its potential applications. Our analysis
remains at a very schematic and theoretical level, and practical
implementations have not been considered. The direction from which 
practical implementations of our scheme and similar schemes may be approached
is uncertain at present. There is, however, a number of possible
practical implementations involving entangled photons \cite{zein,kw},
entangled ions in a linear trap \cite{cir} and entangled nuclear 
spins in NMR \cite{nmr}. In general, any medium that proves to be
useful for quantum computation will immediately  support implementation
of multiparticle entanglement manipulations. This may also enable us
to study the problem of quantifying the amount of entanglement in a given 
multi-particle entangled state \cite{vedral}. For example, we showed how to create an $N$ particle entangled state given an $M$ particle entangled state. 
By studying creation and destruction of entanglement in these measurements we
can perhaps relate amounts of entanglement present in states involving different number of particles. In this paper we have dealt solely with pure states. It should be interesting to generalize our scheme to
arbitrary density matrices. 
   
    A novel feature of entanglement swapping is that it allows superluminal 
establishment of entanglement between two distant particles. This contrasts
with standard schemes of
setting up entanglement which rely on generating entangled particles
at a point and supplying them to distant users and as such can take place
at most at the speed light takes to travel from the source of the
particles to either of the users. Another feature of entanglement swapping is that the particles which are entangled
finally can come from totally independent initial sets of entangled particles,
which means they must have had different values of local hidden variables (if one supposed that they exist). It should then be interesting to investigate whether these
can lead to any stronger type or different type of violation of local hidden variable (LHV) models than that given by Bell's inequality.

\section*{Acknowledgments}
We thank Artur Ekert, Lucien Hardy, Richard Jozsa, Martin Plenio, Sandu Popescu,
Harald Weinfurter and Anton Zeilinger for very useful feedback on earlier versions
of this work. This work was supported in part by the UK Engineering and Physical Sciences Research Council, the European Union, the Inlaks Foundation and by the Knight Trust.

\end{multicols}



\newpage
\vspace{2cm}

\vspace{1cm}
\newpage

\newpage

\newpage

\end{document}